\documentclass[pre,twocolumn,showpacs,superscriptaddress,preprintnumbers,floatfix]{revtex4-1}

\usepackage{etex}
\usepackage{ifpdf}
\usepackage{hyperref}
\usepackage{dcolumn}
\usepackage{url}
\usepackage{amsmath}
\usepackage{amscd}
\usepackage{amsfonts}
\usepackage{amssymb}
\usepackage{amsthm}
\usepackage{xfrac}
\usepackage{braket}
\usepackage{bm}   
\usepackage{bbm}
\usepackage{verbatim}
\usepackage{stmaryrd}
\usepackage{xcolor}
\usepackage{setspace}

\newcommand{\eM}{$\epsilon$-machine}
\newcommand{\eMs}{$\epsilon$-machines}

\def\clap#1{\hbox to 0pt{\hss#1\hss}}

\newenvironment{itemize*}%
  {\begin{itemize}%
    \setlength{\itemsep}{0pt}%
    \setlength{\parskip}{0pt}%
    \setlength{\topsep}{0pt}%
    \setlength{\partopsep}{0pt}%
    \setlength{\parsep}{0pt}}%
  {\end{itemize}}

\newenvironment{enumerate*}%
  {\begin{enumerate}%
    \setlength{\itemsep}{0pt}%
    \setlength{\parskip}{0pt}%
    \setlength{\topsep}{0pt}%
    \setlength{\partopsep}{0pt}%
    \setlength{\parsep}{0pt}}%
  {\end{enumerate}}

\begin{document}

\title{Understanding and Designing Complex Systems:\\
\vspace{0.1in}
Response to\\
``A framework for optimal high-level descriptions\\
in science and engineering---preliminary report''}

\author{James P. Crutchfield}
\email{chaos@ucdavis.edu}
\author{Ryan G. James}
\email{rgjames@ucdavis.edu}
\affiliation{Complexity Sciences Center and Department of Physics,\\
University of California at Davis, One Shields Avenue, Davis, CA 95616}

\author{Sarah Marzen}
\email{smarzen@berkeley.edu}
\affiliation{Department of Physics and
Redwood Center for Theoretical Neuroscience\\
University of California at Berkeley, Berkeley, CA 94720-5800}

\author{Dowman P. Varn}
\email{dpv@complexmatter.org}
\affiliation{Complexity Sciences Center and Department of Physics,\\
University of California at Davis, One Shields Avenue, Davis, CA 95616}

\date{\today}
\bibliographystyle{unsrt}

\begin{abstract}
We recount recent history behind building compact models of nonlinear, complex
processes and identifying their relevant macroscopic patterns or
``macrostates''. We give a synopsis of computational mechanics, predictive
rate-distortion theory, and the role of information measures in monitoring model
complexity and predictive performance. Computational mechanics provides a
method to extract the optimal minimal predictive model for a given process.
Rate-distortion theory provides methods for systematically approximating such
models. We end by commenting on future prospects for developing a general
framework that automatically discovers optimal compact models.  As a response
to the manuscript cited in the title above, this brief commentary corrects
potentially misleading claims about its state space compression method and
places it in a broader historical setting.

\vspace{0.2in}

\end{abstract}

\preprint{Santa Fe Institute Working Paper 14-XX-XXX}
\preprint{arxiv.org:14XX.XXXX [cond-mat.stat-mech]}

\maketitle


\setstretch{1.1}

\setlength{\parskip}{5pt}
\setlength{\parindent}{0pt}

\section{Introduction}

Building compact models of nonlinear processes goes to the heart of our
understanding the complex world around us---a world replete with unanticipated,
emergent patterns. Via discovery mechanisms that we do not yet understand well,
we eventually do come to know many of these patterns, even if we have never
seen them before. Such discoveries can be substantial. At a minimum, compact
models that capture such emergent ``macrostates'' are essential tools in
harnessing complex processes to useful ends. Most ambitiously, one would hope
to automate the discovery process itself, providing an especially useful tool
for the era of Big Data.

One key problem in the larger endeavor of pattern discovery is \emph{dimension
reduction}: reduce the high-dimensional state space of a stochastic dynamical
system into smaller, more manageable models that nonetheless still capture the
relevant dynamics. The study of complex systems always requires this. For
better or worse, it is frequently accomplished in an \emph{ad hoc} fashion
\cite{Gorb06a,Shil08a}. Indeed, it is desirable to have an overarching framework for
this kind of analysis that can be applied across the many manifestations of
complex systems, but to date such a broad theory has not been forthcoming.
Thus, the need for this kind of research remains and is more timely than ever
\cite{Crut09c}. Not surprisingly, it has a long and active history.

This is the setting into which steps a recent \texttt{arxiv.org} preprint ``A
framework for optimal high-level descriptions in science and
engineering---preliminary report"~\cite{Wolp14a}. As a solution to the problem
of dimension reduction, it advocates for \emph{state space compression} (SSC):
Form a compressed variable $Y_t$ that predicts a target statistic $\Omega_t$ of
a system's behavior $\ldots X_{t-1} X_{t} X_{t+1} \ldots$.  When viewed in a
historical context, it is unclear if SSC is more than an alternative notation
and vocabulary for extant approaches to dimension reduction. Here, we explain
this question. We are concerned about several instances in Ref.~\cite{Wolp14a}
where statements are made about research we either participated in or are quite
familiar with that do not accurately reflect that work. The following comments
air our concerns, providing several constructive suggestions.

Our response to Ref. \cite{Wolp14a} is organized as follows. We recall the
history over the last half century driving interest and research on
reconstructing optimal minimal models, specifically as it bears on nonlinear
complex systems. We briefly recount the approach of computational mechanics,
which defined what optimal predictive models are and gave the general solution
to finding them. We then draw connections to predictive rate-distortion theory
that systematically approximates those optimal models. We also comment on
information-theoretic ways to quantify model complexity and predictive
performance and how they trade-off against each other. Our goal is to respond
directly and briefly to Ref. \cite{Wolp14a}, but not to review the broad and
extensive literature on the topic of optimal descriptions of complex systems.
As such, citations are intentionally narrowed to support a single narrative
thread.

\section{Reconstructing Low-Dimensional Models of Complex Processes}

The research program to automate theory building for complex systems has its
origins in fluid turbulence---a high-dimensional system, if there ever was
one---studied for many decades, of great practical import, and at the time,
according to Heisenberg \cite{Heis67a}, one of the premier problems in
nonlinear physics. Cracking this problem relied on developing a connection
between the abstract theory of dynamical systems and hydrodynamic experiment.
This came in the \emph{attractor reconstruction} method that extracted
``Geometry from a Time Series'' \cite{Pack80} using one or a few signals from a
high-dimensional complex system. Attractor reconstruction eventually led to
demonstrating that deterministic chaos is the mechanism generating weak
turbulence \cite{Bran83}, verifying a long-standing conjecture \cite{Ruel71a}
in dynamical systems theory that over-threw the decades-old quasi-periodic
theory of turbulence due to Landau and Lifschitz \cite{Land59a}.

The reconstruction method, though, only produced a reduced-dimension space of
\emph{effective states} of the infinite-dimensional fluid dynamics, ignoring
the dynamical mechanism that generated the turbulent motion. Generalizing
attractor reconstruction, Refs. \cite{Crut87a,Farm87} introduced methods to
infer the \emph{effective theory} or equations of motion from time-series
generated by chaotic dynamical systems. Reference \cite{Crut87a}, in
particular, also provided a critique of the general reconstruction approach,
highlighting its subjectivity---one must choose a class of model
representation. Such \emph{ad hoc} choices preclude an objective
measure of a system's complexity. Reference \cite{Crut88a} solved the
subjectivity problem. It was the first to state the problem of optimal
predictive, \emph{minimal} models of complex processes and provided its
solution---the \emph{\eM} and its \emph{causal states}. Using this foundation,
it was able to define the \emph{statistical complexity}---the first consistent,
constructive measure of structural complexity. And, it
introduced \emph{intrinsic computation}---the idea that all physical systems
store and process information. Influenced by the goals of artificial
intelligence at the time, it also challenged future researchers to develop an
\emph{artificial science}---automating the construction of minimal causal
models of complex systems. Using these methods, which came to be called
\emph{computational mechanics}, Ref. \cite{Crut92c} went on to give the first
quantitative definition of the emergence of macrostate organization in terms of
increasing structural complexity and intrinsic computation. It argued that
there is a natural hierarchy of causal organization and introduced a
renormalization-group method for moving up the hierarchy---a method of genuine
pattern discovery. A recent review of this history is found in Ref.
\cite{Crut12a}.

\section{Causal States and Macrostates}

Identifying emergent organization, especially if pursued as a problem in
theoretical physics, is often couched in terms of finding system
\emph{macrostates}. The metaphorical intuition behind this framing is roughly
that emergent organization is the analog of the macroscopic, measurable
properties of a thermodynamic system. A macrostate---say, given by particular
values of temperature and pressure---is the set of the many microscopic
molecular configurations or \emph{microstates} that lead to the same measurable
properties. In this framing, macrostates emerge from the microstates under the
action of the dynamics on the microscopic scale as it relaxes to equilibrium
\cite{Cros93a}. The practicing statistical physicist, more specifically, often
begins the analysis of a system's properties by searching for an ``order
parameter'' or for insightful ``coarse-grainings'', which are analogous
concepts at our general level of discussion here.

A system's causal states, being groupings of microscopic trajectories that
capture a system's emergent behaviors and organization, play a role very
analogous to the system's macrostates and not its microstates \footnote{We
describe them as \emph{mesoscopic states}---a structural level intermediate
between raw system configurations (microstates) and the macrostates of
thermodynamics.}. Reference \cite{Wolp14a} offers a view that is substantially
at variance with this. SSC addresses this in terms of the descriptions to which
given behavior is compressed. Notably, Ref.~\cite[page 6]{Wolp14a} starts by
restricting the microscopic dynamics:
\begin{quote}
The \textbf{microstate} ... evolves in time according to a (usually) Markovian process \ldots
\end{quote}
And then, on page 41 it notes that the observations of the stochastic processes
analyzed in computational mechanics:
\begin{quote}
\dots do not evolve according to a first-order Markov process, and so cannot be
identified with the fine-grained values $X_t$ of SSC. On the other hand, \dots,
evolution over the space of causal sets is first-order Markov. This suggests
that we identify the causal states with SSC's fine-grained space, not its
compressed space.
\end{quote}
In the page 6 quote, however, SSC assumed first-order Markov dynamics on the
underlying process (microstates). In the context described above, in which the
target $\Omega_t$ is the future $X_t X_{t+1} \ldots $ and $Y_t$ compresses the
past $\ldots X_{t-2} X_{t-1}$, the causal states are interpreted as
coarse-grainings \emph{only} of $X_t$. Thus, in SSC the dynamics on the causal
states and the observed process are \emph{both first-order Markov}. These
restrictions are unnecessary.

A telling consequence of SSC's misidentification of how computational mechanics
is used is that the assumption of a first-order Markov process for the
microstates simplifies the dimension-reduction problem to the point that the
computational complexity of identifying causal states would fall in P and no
longer be NP-hard, as it is more generally. Overall, the simplification to
first-order Markov obscures the relationship between computational mechanics
and SSC.  Computational mechanics can be applied to any stochastic process,
including the Markovian one assumed to govern SSC's microscopic variables
$\ldots X_{t-1} X_t X_{t+1} \ldots$.

Apparently, despite being defined as a coarse-graining, SSC (incorrectly)
associates causal states with its microstates $X_t$ for no reason other than
their Markovian nature. This is a confusing association for two reasons: First,
the $X_t$ values were introduced as being (usually) Markovian, not necessarily
Markovian; and second, by construction computational mechanics' causal states
are a coarse-graining of trajectories and so by definition are a form of
state-space compression. SSC's association is based on reasoning that is
fundamentally flawed, reflecting a basic misunderstanding of causal states and
how to apply computational mechanics.

\section{State Space Compression versus Rate Distortion Theory}

Now, let's turn to consider how SSC defines its ``macrostates'' via
coarse-graining microstates. In this, we find a basic connection between SSC
and Shannon's rate-distortion theory applied to prediction---what is called
\emph{predictive rate distortion} (PRD) \cite{Marz14f}. SSC largely ignores
this important connection to prior work. This is strange since Shannon
introduced rate-distortion theory explicitly to solve dimension reduction
problems \cite{Shan48a,Shan59a}: find a compact ``encoding'' of a data source
subject to a set of constraints, such as transmission rate, accuracy,
processing, delay, and the like. Optimal modeling can be framed in just this
way, for example, as found in Rissanen's \emph{minimum description length}
approach \cite{Riss86a}. The physics metaphor for building models extends to
this setting, too: data are the microstates, compressed model variables are
macrostates, and coding constraints are physical boundary conditions.

Take SSC's target $\Omega_t$ to be the future of $X_t$, and find some $Y_t$ that
compresses the past of $X_t$ to retain expected accuracy in predicting
$\Omega_t$ without unnecessarily increasing asymptotic coding cost of
transmitting $Y_t$ to another observer. When accuracy is heavily prioritized
over coding cost, the causal states are recovered, as shown by Refs.
\cite{Stil07a,Stil07b} and as discussed in Ref.~\cite{Stil14a}. In fact, the
causal states are essentially the answer to the question: Among all possible
compressions with minimal accuracy cost, which has the minimal coding cost?
Both issues of accuracy and coding cost can be extended directly to the case in
which the target $\Omega_t$ is some coarse-graining of the future of $X_t$, so
that one searches for compressed predictive and perceptual features or
macrostates. Recent PRD work shows, in fact, how to extract just such
coarse-grained macrostates, given a process's \eM\ \cite{Marz14f}, noting that
the latter can be calculated theoretically \cite{Crut97a,Feld98b} or estimated
empirically \cite{Shal02a,Stre13a,Varn14a}.

However, a major conceptual difference between PRD and the SSC framework is
that SSC explicitly restricts $X_t$ to be first-order Markov; whereas, PRD can
address arbitrary order and infinite-order Markov processes. The first-order
Markov assumption is simply not realistic. Microstates are rarely
experimentally accessible; e.g., the spike trains from neurophysiological
studies are a very coarse-grained observable of underlying membrane voltage
fluctuations \cite{Riek99} and, as such, their dynamics is often infinite-order
Markov \cite{Marz14e}. This remains true even if the underlying membrane
voltage dynamics are first-order Markov.

PRD and computational mechanics actually do state space compression. They
find compressed predictive representations of a time series. Moreover,
computational mechanics finds a hidden Markov model (HMM) for the given
(non-Markovian) time series. Indeed, that's the point of the \eM\ and, for
lossy versions, the point of causal rate-distortion theory \cite{Marz14f} and
the recursive information bottleneck method \cite{Stil14a}.

\section{Coding Cost}

This all leads to the central question of how to quantify the organization
captured by these macrostates. PRD and computational mechanics use the
statistical complexity---the Shannon information in the causal states or in
their coarse-grainings. SSC takes issue with the use of information theory in
these approaches \cite[page 40]{Wolp14a}:
\begin{quote}
Statistical complexity plays a role in the objective function of [77] that is
\emph{loosely analogous} to the role played by accuracy cost in the SSC objective function.
\end{quote}
[Emphasis ours.]
However, SSC's coding cost $H[Y_0]$ is the information contained in its version
of the causal states. This is a simplified version of the statistical
complexity $C_\mu$, once one restricts to first-order Markov processes and
predictive mappings from pasts $\ldots X_{t-2} X_{t-1}$ to $Y_t$. That is,
$C_\mu$ is a \emph{coding cost}, not an accuracy cost.

Moreover, on Ref. \cite[page 40]{Wolp14a}, we read:
\begin{quote}
\dots the authors consider a ratio of costs rather than a linear combination of them, like we consider here.
\end{quote}
However, PRD objective functions---e.g., as described by
Refs.~\cite{Shal99a,Stil14a}---are linear combinations of mutual informations.

\section{Mutual Information as Accuracy Cost}

Reference \cite[page 46]{Wolp14a} criticizes the use of general multivariate
mutual information in noting a difficulty with systems governed by a
time-varying dynamic:
\begin{quote}
The underlying difficulty is inherent in the very notion of using mutual
information to define accuracy cost, and is intrinsic to consideration of the
relation between $Y$ and $X$ at more than two moments in time.
\end{quote}
Using mutual information at more than two moments in time is simply not a
problem for ergodic stationary processes \cite{Jame11a}. Moreover, it's not
necessarily a problem for nonergodic or nonstationary processes either. For
these, several examples in Ref. \cite{Bial01a}, as well as those in
Refs.~\cite{Trav11b,Tche13a}, calculate the past-future mutual information
(\emph{excess entropy}) that involves all moments in time.

However, there is perhaps another underlying difficulty with using mutual
information as an accuracy cost: It is used alternatively as either the coding
cost or an inverse accuracy cost in the information bottleneck method
\cite{Bial00a}. And, there, mutual information as an `inverse accuracy cost'
really amounts to an accuracy cost that employs the KL-divergence between
$\Pr(\Omega_{t} |Y_t)$ and $\Pr(\Omega_{t} | X_t)$ in the more general PRD
framework.

Generally, though, the rate-distortion theorem \cite{Cove06a} justifies the use
of mutual information as a coding cost regardless of distortion measure. And,
helpfully, this has been extended to the nonasymptotically large block coding
limit \cite{Kost12a,Kost13a}, demonstrating that one should not be glib about
introducing new coding costs.

\section{Generality}

SSC is offered as an improvement on the current literature for being a
principled and constructive method of generating macrostates. Reference
\cite{Wolp14a}'s abstract states:
\begin{quote}
This State Space Compression framework makes it possible to solve for the
optimal high-level description of a given dynamical system $\ldots$.
\end{quote}
This brings up two concerns, one explicit in the quote and one implicit.

First, what is provided is not a constructive framework; it does not provide
methods to solve for the macrostates. Moreover, each of the provided macrostate examples is constructed in an \emph{ad hoc} manner.

Second, the burden of proof lies with SSC's advocates. Since alternative
frameworks exist and are contrasted against, at this date progress behooves the
authors to provide examples where their framework succeeds and others fail.

SCC's lack in these regards should be compared with how alternatives had to
develop new calculational methods for the challenging problems that are
entailed in modeling complex systems. For example, computational mechanics
extended methods from holomorphic functional calculus that now give
closed-form expressions for a process's information measures \cite{Crut13a}.

We emphasize that SSC's $H(Y_0)$ is exactly the computational cost-based
objective function used to identify causal states, in which we constrain
ourselves to deterministic mappings $Y_t = f(X_t)$ such that
$\Pr(\Omega_t|Y_{t_0}) = \Pr(\Omega_t|f(X_{t_0}))$. Indeed, more broadly
interpreted, causal states are the unique and minimal macrostate choice for SSC
in the limit of high premium on accuracy and minimal concern about
computational cost. Adapting the proof of Thm. 1 in Ref.~\cite{Stil07b} will be
helpful here. In any case, the causal states are the minimal sufficient
statistic for prediction \cite{Shal98a}. In other words, any process statistic
can be calculated from them and this raises the bar quite high for SSC's
proposed alternative macrostates.

In terms of implementations, PRD and computational mechanics are constructive.
In rate-distortion theory generally there's the Blahut-Arimoto algorithm and its generalizations for calculating coarse-grained macrostates \cite{Rose94a,Cove06a}. And, there are several statistical inference algorithms that estimate \eMs\ from various kinds of data \cite{Shal02a,Pfau11a,Varn12a,Stre13a}.  

The general problem of dimension reduction for complex systems is an important
one, and we encourage efforts along these lines. We appreciate that the
synopses of computational mechanics, PRD, and information measures above
address but a small part SSC's stated goals and the goals earlier researchers
have set. It is important, though, that SSC start with correct assumptions and
an understanding of its antecedents. In any case, we hope that our comments
remedy, in a constructive way, misleading impressions that Ref. \cite{Wolp14a}
gives of the state of the art of computational mechanics and predictive rate
distortion.

\vspace{-0.2in}
\section{Rubber, Meet Road}
\vspace{-0.1in}

Finally, let's end on a practical note. We are advocates for a broad, even
pluralistic approach to automated nonlinear model building---i.e., for
artificial science. However, our repeated experience is that general
``frameworks'' seriously stub their toes on application to experiments. Despite
this, we are still optimists. Those wishing to contribute, however, should pick
at least one application area and drill all the way down to show their
alternative discovers a particular new scientific phenomenon. This is a
necessary calibrating exercise for any attempt at generality.

In contrast to SSC, its antecedents have done their due diligence.
Rate-distortion theory, developed for over a decade into the information
bottleneck method \cite{Bial00a}, has been applied to test
how close sensory spike trains are to stimuli predictive
information functions \cite{Palm13a}. For our part,
computational mechanics led to a new theory of structure in disordered
materials \cite{Varn14a} and to measuring novel information processing in
neural spike trains \cite{Marz14e}. Looking forward to future engineering of
complex systems, the structural understanding developed in these applications
moves us in a direction to design novel semiconducting materials for a new
generation of computing technology and the next generation of nanoscale
spike-train probes that will scale to monitoring the activity of thousands of
neurons \cite{Aliv12a}.

\vspace{-0.1in}
\acknowledgments
\vspace{-0.1in}

The authors thank the Santa Fe Institute for its hospitality during visits and
Chris Ellison, Dave Feldman, and John Mahoney for helpful comments. JPC is an
SFI External Faculty member. This material is based upon work supported by, or
in part by, the U.S. Army Research Laboratory and the U. S. Army Research
Office under contracts W911NF-13-1-0390, W911NF-13-1-0340, and
W911NF-12-1-0288. S.M. is a National Science Foundation Graduate Student
Research Fellow and a U.C. Berkeley Chancellor's Fellow.

\bibliography{chaos}

\begin{thebibliography}{10}

\bibitem{Gorb06a}
A.~N. Gorban, N.~Kazantzis, I.~G. Kevrekidis, H.~C. Ottinger, and
  C.~Theodoropoulos, editors.
\newblock {\em Model Reduction and Coarse-Graining Approaches for Multiscale
  Phenomena}.
\newblock Series in Complexity. Springer, 2006.

\bibitem{Shil08a}
W.~H.~A. Schilders, H.~A. van~der Vorst, and J.~Rommes.
\newblock {\em Model Order Reduction: {Theory}, Research Aspects and
  Applications}, volume~13 of {\em Mathematics in Industry}.
\newblock Springer, 2008.

\bibitem{Crut09c}
J.~P. Crutchfield.
\newblock The dreams of theory.
\newblock {\em WIRES Comp. Stat.}, 6(March/April):75--79, 2014.

\bibitem{Wolp14a}
D.~H. Wolpert, J.~A. Grochow, E.~Libby, and S.~De{D}eo.
\newblock A framework for optimal high-level descriptions in science and
  engineering---{Preliminary} report, 2014.
\newblock arxiv.org: 1409.7403.

\bibitem{Heis67a}
W.~Heisenberg.
\newblock Nonlinear problems in physics.
\newblock {\em Physics Today}, 20:23--33, 1967.

\bibitem{Pack80}
N.~H. Packard, J.~P. Crutchfield, J.~D. Farmer, and R.~S. Shaw.
\newblock Geometry from a time series.
\newblock {\em Phys. Rev. Let.}, 45:712, 1980.

\bibitem{Bran83}
A.~Brandstater, J.~Swift, Harry~L. Swinney, A.~Wolf, J.~D. Farmer, E.~Jen, and
  J.~P. Crutchfield.
\newblock Low-dimensional chaos in a hydrodynamic system.
\newblock {\em Phys. Rev. Lett.}, 51:1442, 1983.

\bibitem{Ruel71a}
D.~Ruelle and F.~Takens.
\newblock On the nature of turbulence.
\newblock {\em Comm. Math. Phys.}, 20:167--192, 1971.

\bibitem{Land59a}
L.~D. Landau and E.~M. Lifshitz, editors.
\newblock {\em Fluid Mechanics}.
\newblock Pergamon Press, Oxford, United Kingdom, 1959.

\bibitem{Crut87a}
J.~P. Crutchfield and B.~S. McNamara.
\newblock Equations of motion from a data series.
\newblock {\em Complex Systems}, 1:417 -- 452, 1987.

\bibitem{Farm87}
J.~D. Farmer and J.~Sidorowitch.
\newblock Predicting chaotic time series.
\newblock {\em Phys. Rev. Lett.}, 59:366, 1987.

\bibitem{Crut88a}
J.~P. Crutchfield and K.~Young.
\newblock Inferring statistical complexity.
\newblock {\em Phys. Rev. Let.}, 63:105--108, 1989.

\bibitem{Crut92c}
J.~P. Crutchfield.
\newblock The calculi of emergence: Computation, dynamics, and induction.
\newblock {\em Physica D}, 75:11--54, 1994.

\bibitem{Crut12a}
J.~P. Crutchfield.
\newblock Between order and chaos.
\newblock {\em Nature Physics}, 8(January):17--24, 2012.

\bibitem{Cros93a}
M.~C. Cross and P.~C. Hohenberg.
\newblock Pattern formation outside of equilibrium.
\newblock {\em Rev. Mod. Phys.}, 65(3):851--1112, 1993.

\bibitem{Note1}
We describe them as \protect \emph {mesoscopic states}---a structural level
  intermediate between raw system configurations (microstates) and the
  macrostates of thermodynamics.

\bibitem{Marz14f}
S.~Marzen and J.~P. Crutchfield.
\newblock Circumventing the curse of dimensionality in prediction: {Causal}
  rate-distortion for infinite-order markov processes.
\newblock 2014.
\newblock SFI Working Paper 14-12-047; arxiv.org:1412.2859
  [cond-mat.stat-mech].

\bibitem{Shan48a}
C.~E. Shannon.
\newblock A mathematical theory of communication.
\newblock {\em Bell Sys. Tech. J.}, 27:379--423, 623--656, 1948.

\bibitem{Shan59a}
C.~E. Shannon.
\newblock Coding theorems for a discrete source with a fidelity criterion.
\newblock {\em IRE National Convention Record, Part 4}, 7:142--163, 623--656,
  1959.

\bibitem{Riss86a}
J.~Rissanen.
\newblock Stochastic complexity and modeling.
\newblock {\em Ann. Statistics}, 14:1080, 1986.

\bibitem{Stil07a}
S.~Still and J.~P. Crutchfield.
\newblock Structure or noise?
\newblock 2007.
\newblock Santa Fe Institute Working Paper 2007-08-020; arxiv.org
  physics.gen-ph/0708.0654.

\bibitem{Stil07b}
S.~Still, J.~P. Crutchfield, and C.~J. Ellison.
\newblock Optimal causal inference: {Estimating} stored information and
  approximating causal architecture.
\newblock {\em CHAOS}, 20(3):037111, 2010.

\bibitem{Stil14a}
S.~Still.
\newblock Information bottleneck approach to predictive inference.
\newblock {\em Entropy}, 16:968--989, 2014.

\bibitem{Crut97a}
J.~P. Crutchfield and D.~P. Feldman.
\newblock Statistical complexity of simple one-dimensional spin systems.
\newblock {\em Phys. Rev. E}, 55(2):R1239--R1243, 1997.

\bibitem{Feld98b}
D.~P. Feldman and J.~P. Crutchfield.
\newblock Discovering non-critical organization: {S}tatistical mechanical,
  information theoretic, and computational views of patterns in simple
  one-dimensional spin systems.
\newblock 1998.
\newblock {Santa} Fe Institute Working Paper 98-04-026.

\bibitem{Shal02a}
C.~R. Shalizi, K.~L. Shalizi, and J.~P. Crutchfield.
\newblock Pattern discovery in time series, {Part} {I}: Theory, algorithm,
  analysis, and convergence.
\newblock 2002.
\newblock Santa Fe Institute Working Paper 02-10-060;
  arXiv.org/abs/cs.LG/0210025.

\bibitem{Stre13a}
C.~C. Strelioff and J.~P. Crutchfield.
\newblock Bayesian structural inference for hidden processes.
\newblock {\em Phys. Rev. E}, 89:042119, 2014.
\newblock Santa Fe Institute Working Paper 13-09-027, arXiv:1309.1392
  [stat.ML].

\bibitem{Varn14a}
D.~P. Varn and J.~P. Crutchfield.
\newblock Chaotic crystallography: {How} the physics of information reveals
  structural order in materials.
\newblock {\em Curr. Opin. Chem. Eng.}, 7:47--56, 2015.
\newblock Santa Fe Institute Working Paper 14-09-036; arxiv.org:1409.5930
  [cond-mat].

\bibitem{Riek99}
F.~Rieke, D.~Warland, R.~de~Ruyter~van Steveninck, and W.~Bialek.
\newblock {\em Spikes: {Exploring} the Neural Code}.
\newblock Bradford Book, New York, 1999.

\bibitem{Marz14e}
S.~Marzen and J.~P. Crutchfield.
\newblock Time resolution dependence of continuous-time information measures.
\newblock {\em in preparation}, 2014.

\bibitem{Shal99a}
C.~R. Shalizi and J.~P. Crutchfield.
\newblock Information bottlenecks, causal states, and statistical relevance
  bases: {How} to represent relevant information in memoryless transduction.
\newblock {\em Adv. Compl. Sys.}, 5(1):91--95, 2002.

\bibitem{Jame11a}
R.~G. James, C.~J. Ellison, and J.~P. Crutchfield.
\newblock Anatomy of a bit: {Information} in a time series observation.
\newblock {\em CHAOS}, 21(3):037109, 2011.

\bibitem{Bial01a}
W.~Bialek, I.~Nemenman, and N.~Tishby.
\newblock Complexity through nonextensivity.
\newblock {\em Physica A}, 302:89--99, 2001.

\bibitem{Trav11b}
N.~Travers and J.~P. Crutchfield.
\newblock Infinite excess entropy processes with countable-state generators.
\newblock {\em Entropy}, 16:1396--1413, 2014.
\newblock SFI Working Paper 11-11-052; arxiv.org:1111.3393 [math.IT].

\bibitem{Tche13a}
M.~Tchernookov and I.~Nemenman.
\newblock Predictive information in a nonequilibrium critical model.
\newblock {\em J. Stat. Phys.}, 153:442--459, 2013.

\bibitem{Bial00a}
W.~Bialek, I.~Nemenman, and N.~Tishby.
\newblock Predictability, complexity, and learning.
\newblock {\em Neural Computation}, 13:2409--2463, 2001.

\bibitem{Cove06a}
T.~M. Cover and J.~A. Thomas.
\newblock {\em Elements of Information Theory}.
\newblock Wiley-Interscience, New York, second edition, 2006.

\bibitem{Kost12a}
V.~Kostina and S.~Verdu.
\newblock Fixed-length lossy compression in the finite blocklength regime.
\newblock {\em IEEE Trans. Info. Th.}, 58(26):3309--3338, 2012.

\bibitem{Kost13a}
V.~Kostina and S.~Verdu.
\newblock Lossy joint source-channel coding in the finite blocklength regime.
\newblock {\em IEEE Trans. Info. Th.}, 59(5):2545--2574, 2013.

\bibitem{Crut13a}
J.~P. Crutchfield, P.~Riechers, and C.~J. Ellison.
\newblock Exact complexity: {Spectral} decomposition of intrinsic computation.
\newblock submitted.
\newblock Santa Fe Institute Working Paper 13-09-028; arXiv:1309.3792
  [cond-mat.stat-mech].

\bibitem{Shal98a}
C.~R. Shalizi and J.~P. Crutchfield.
\newblock Computational mechanics: Pattern and prediction, structure and
  simplicity.
\newblock {\em J. Stat. Phys.}, 104:817--879, 2001.

\bibitem{Rose94a}
K.~Rose.
\newblock A mapping approach to rate-distortion computation and analysis.
\newblock {\em IEEE Trans. Info. Th.}, 40(6):1939--1952, 1994.

\bibitem{Pfau11a}
D.~Pfau, N.~Bartlett, and F.~Wood.
\newblock Probabilistic deterministic infinite automata.
\newblock In {\em Adv. Neural Info. Proc. Sys.}, pages 1930--1938. MIT Press,
  2011.

\bibitem{Varn12a}
D.~P. Varn, G.~S. Canright, and J.~P. Crutchfield.
\newblock {$\epsilon$-Machine} spectral reconstruction theory: {A} direct
  method for inferring planar disorder and structure from {X}-ray diffraction
  studies.
\newblock {\em Acta. Cryst. Sec. A}, 69(2):197--206, 2013.

\bibitem{Palm13a}
S.~E. Palmer, O.~Marre, M.~J.~Berry II, and W.~Bialek.
\newblock Predictive information in a sensory population.
\newblock 2013.
\newblock arXiv:1307.0225.

\bibitem{Aliv12a}
A.~P. Alivisatos, M.~Chun, G.~M. Church, R.~J. Greenspan, M.~L. Roukes, and
  R.~Yuste.
\newblock The brain activity map project and the challenge of functional
  connectomics.
\newblock {\em Neuron}, 74:970--974, 2012.

\end{thebibliography}

\end{document}